\documentclass[prb, twocolumn]{revtex4}
\usepackage[dvips]{graphics}

\usepackage{amsthm}
\usepackage{graphicx}
\usepackage{oldgerm}
\usepackage{latexsym}
\usepackage{amsmath, mathrsfs, amssymb, amscd}
\usepackage{longtable}

\newcommand{\eps}{\varepsilon_1}  
\newcommand{\sig}{\sigma_1}       
\newcommand{\e}{\varepsilon}

\newcommand{\rhop}{r_\omega}     
\newcommand{\WW}{U(\rhop)}       

\begin{document} 
\title{Analysis of broadband microwave conductivity and permittivity measurements of semiconducting materials}

\author{Elvira Ritz}
\author{Martin Dressel}
\affiliation{1.~Physikalisches Institut, Universit{\"a}t
Stuttgart, Pfaffenwaldring 57, D-70550 Stuttgart, Germany}

\date{\today}
\begin{abstract}
We perform broadband phase sensitive measurements of the
reflection coefficient from 45 MHz up to 20 GHz employing a vector
network analyzer with a 2.4 mm coaxial sensor which is terminated
by the sample under test. While the material parameters
(conductivity and permittivity) can be easily extracted from the
obtained impedance data if the sample is metallic, no direct
solution is possible if the material under investigation is an
insulator. Focusing on doped semiconductors with largely varying
conductivity, here we present a closed calibration and evaluation
procedure for frequencies up to 5 GHz, based on the rigorous
solution for the electromagnetic field distribution inside the
sample combined with the variational principle; basically no
limiting assumptions are necessary. A simple static model based on
the electric current distribution proves to yield the same
frequency dependence of the complex conductivity up to 1 GHz.
After a critical discussion we apply the developed method to the
hopping transport in Si:P at temperature down to 1 K.
\end{abstract}


\maketitle %

\newpage

\section{Introduction} The rapid development of communication as
well as industrial and medicine technologies demands accurate
characterization of components at ever increasing frequencies,
i.e.\ beyond the radio frequency range; this becomes in particular
relevant for insulating and semiconducting materials employed in
electronic devices\cite{Shur}.  Beside the high technological
relevance of doped semiconductors they are also subject to intense
fundamental research concerning disordered electronic systems with
electron-electron correlations \cite{EfPo,Lohneysen98}.

On a macroscopic scale and under steady-state conditions, the
interaction of a material with electric field is determined by its
conductivity and dielectric permittivity. The desired broadband
characterization of those parameters becomes more challenging with
rising frequencies because losses and spatial variation of current
and voltage then gain importance.

The materials characterization up to the MHz range turns out to be
comparably simple: the voltage drop is measured when a current
passes homogeneously through the specimen; lock-in technique
allows for the determination of the complex response. As soon as
the GHz range is approached, the wavelength becomes comparable to
the leads and specimen, waveguides have to be utilized and
reflection or transmission coefficients are measured. In this
spectral range a vector network analyzer is a suited and powerful
tool.

While standard circuit theory applies to radio frequencies, in the
microwave range the wavelength becomes as short as a few
millimeters and thus a careful treatment of the electromagnetic
field distribution within the sample is necessary to obtain the
material parameters from the impedance data gained by the
measurement. Whereas the evaluation is straightforward for
metallic samples under investigation \cite{Drs,Anlage,SD}, only
some approximate solutions and models have been developed in the
past to treat the dielectric materials
\cite{Stuchlies}$^-$\cite{PM}, considering in most cases liquids
or soft matter at ambient conditions. In the course of
investigating the dynamical conductivity of doped semiconductors
at low temperatures, we revisited the existing methods and
elaborated a reliable evaluation procedure with optimized
theoretical and experimental complexity. We are now able to obtain
both the real and imaginary parts of the conductivity
$\sigma=\sig+i\sigma_2$ (or the dielectric function
$\e=\eps+i\e_2$) by a broadband measurement from 0.1~GHz to 5~GHz.
(Throughout this text, $\e$ is the complex dielectric function of
the material, relative to the free space permittivity $\e_0$.)

As an application, the conductivity and permittivity of insulating
Si:P at temperatures down to 1~K in a broad frequency and
donor-concentration range have been investigated. In particular,
those measurements provide new information on the influence of
electronic correlations on the hopping transport in Si:P as well
as on the critical scaling of its dielectric constant at the
metal-insulator transition driven by the doping concentration.

We start in Sec.~\ref{general} with the broadband microwave
measurement of a general solid sample terminating the coaxial
line. Following a brief survey of the well-known evaluation
methods for metallic materials, the problems are formulated which
arise for samples with non-metallic conductivity. A simple static
model is presented in Sec.~\ref{static_model} for the current
distribution inside a semiconducting sample, valid in the
low-frequency range up to 1 GHz. In Sections \ref{LF} and
\ref{open_calibration} it is followed by a rigorous solution to
the problem making no severe simplifications. Finally, the method
is applied to the hopping transport in Si:P in Section
\ref{application}.

\section{Broadband microwave conductivity and
permittivity measurements}\label{general}

\begin{figure}
  \includegraphics*[width=\linewidth]{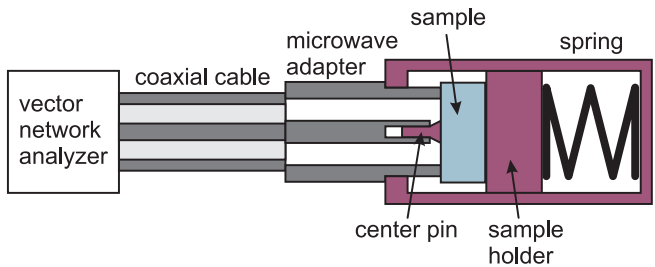}
  \caption{(color online)
Schematics of a broadband microwave spectrometer that employs a
vector network analyzer and coaxial waveguides. A solid sample terminates the otherwise open-ended coaxial line (after
Ref.~\protect\onlinecite{Anlage}).}\label{sceme}
\end{figure}
The experimental arrangement for complex permittivity
measurements, where the material under investigation is placed at
the aperture of a coaxial probe, allows for a broadband phase
sensitive measurement of the reflection coefficient using a vector
network analyzer. In case of solid matter some smart jig is needed
to press the flat sample surface against the sensor
\cite{Anlage,SD}, as depicted in Fig.~\ref{sceme}. Metallic
contacts (gold or aluminium) are usually evaporated on top of the
solid specimen that match the inner and outer conductors of the
coaxial probe in size. These contacts provide a proper electrical
connection between the sample material and the probe, and they
define the geometry of the sample surface exposed to the signal.
The sensor can be inserted into a temperature-controlled
environment and, for a given probe size, there is a useable
frequency range as broad as two orders of magnitude.

The problem of extracting the interesting material parameters from
the measured reflection coefficient data can be solved in two
steps:  first, one needs to obtain the complex sample impedance
from the measured reflection coefficient ($S$ parameter), and
second, the complex material properties have to be calculated from
the impedance.

\subsection{Evaluation procedure for metallic
samples}\label{metallic}

The evaluation of metallic samples has been developed and
experimentally tested in the recent years \cite{Anlage,SD}:\\

The first task is to obtain the complex sample impedance $Z$ from
the reflection coefficient $S_{11,m}$, measured by the test set of
the network analyzer, like the HP 8510. The general error model
for a reflection measurement \cite{HFLange} results in the
following relation:
\begin{eqnarray}
S_{11,m}=E_D+\frac{E_RS_{11}}{1-E_SS_{11}}\, ,
\end{eqnarray}
between the measured $S$ parameter and the actual reflection
coefficient $S_{11}$ of the sample. The three independent complex
values $E_R$, $E_S$ and $E_D$ comprise the contribution of the
microwave line. To determine those, measurements of three
independent calibration samples with known actual reflection
coefficients $S_{11}$ as functions of frequency and temperature
are required. We use bulk aluminium samples as short, teflon
samples as open and thin metallic NiCr films as load standards.
\begin{figure}
\centering
  \includegraphics*[width=\linewidth]{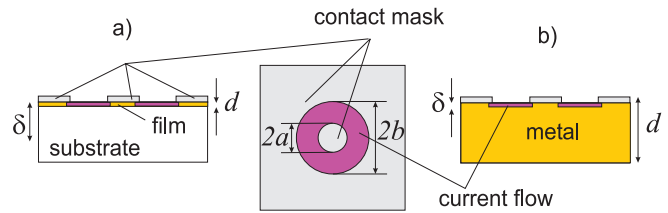}
  \caption{(color online) Current flow (dark magenta area) in a metallic sample (bright yellow) with a contact mask (light grey) on top
of it; in our case $2a=0.6$ mm and $2b=1.75$ mm. The relevant
limiting cases a) $d\ll\delta$ and b) $d\gg\delta$ lead to a
simple evaluation
  procedure ($\delta < 1 \mu$m is the skin depth).}\label{metal}
\end{figure}

The sample impedance $Z$ then follows directly via
\cite{Pozar}:
\begin{eqnarray}
Z= \frac{1+S_{11}}{1-S_{11}}Z_0 \, \label{Z-S} \, ,
\end{eqnarray}
where $Z_0=50~\Omega$ is the characteristic impedance of the
microwave line.\\

The way to extract the complex electric conductivity $\sigma$ of a
metallic sample from its complex impedance $Z$ is straightforward,
if the thickness of the specimen $d$ either significantly exceeds
the skin depth $\delta$ or vice versa:
\begin{itemize}
\item {\bf $d\ll\delta$.} (Fig.~\ref{metal}a) In case of a thin
film evaporated on an insulating substrate, the electric field
strength stays nearly constant throughout the whole film thickness
$d$. The relation between the conductivity $\sigma$ and $Z$
depends only on the geometry of the contacts. For the ring of
inner radius $a$ and outer radius $b$ between the contacts, it
reads:
\begin{eqnarray}
Z=\frac{1}{\sigma}\frac{\ln\{b/a\}}{2\pi d}\, . \label{film}
\end{eqnarray}
\item {\bf $d\gg\delta$.} (Fig.~\ref{metal}b) For typical
microwave frequencies, the electromagnetic wave in a thick
metallic sample is significantly damped already at the depth of
1~$\mu$m by the skin effect. Hence, the interaction with the
incident wave takes place in a thin layer at the sample surface.
Boundary effects at the edges of the relatively broad contact area
(cf.\ Fig.~\ref{metal}) are negligible and the concept of the
surface impedance $Z_S$ based on the assumption of plane wave
propagation works well \cite{Drs}. In case of a ring with inner
and outer radii $a$ and $b$, the formula to extract the
conductivity from the measured impedance $Z$ reads
\cite{simplified}:
\begin{eqnarray}
\sigma&=&\frac{\omega}{i}\left(\frac{\mu_0}{Z_S^2}-\e_0\right)\nonumber \\
&=&\frac{\omega}{i}\left(\frac{\mu_0(\ln\{b/a\})^2}{(2\pi
Z)^2}-\e_0\right)\, ,\label{sigma-Zs}
\end{eqnarray}
with the magnetic permeability of vacuum $\mu_0$ and the free
space permittivity $\e_0$.
\end{itemize}

\subsection{Formulation of the problem for semiconducting
materials} \label{problem} 
\begin{figure}
\centering
\includegraphics*[width=0.75\linewidth]{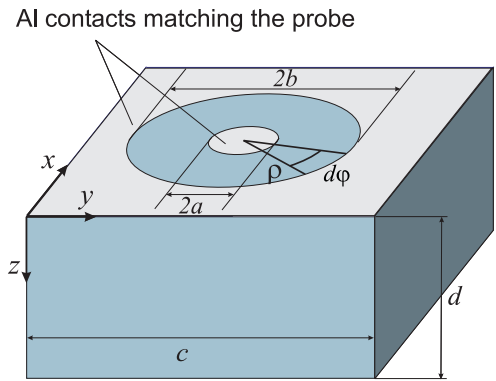}
\caption{(color online) Sketch of a semiconducting sample with the
aluminium contact-layer (light grey) on top. $2a=0.6$ mm,
$2b=1.75$ mm, corresponding to a 2.4 mm microwave adapter, $c=5$
mm, $d=2$ mm}\label{sample}
\end{figure}

For materials with non-metallic
conductivity, none of the simple assumptions valid for metallic
samples hold and both steps of the evaluation procedure contain
additional challenge:
\begin{enumerate}
\item For the calibration of the microwave line the open standard
is as significant as the short and the load standards, when
insulating samples are measured. It acts as a complex capacitor
and the dc assumption $S^{\rm open}_{11}=1$, corresponding to
$Z^{\rm open}=\infty$, is not sufficient at higher frequencies.
The correct frequency dependence of the reflection coefficient
$S^{\rm open}_{11}$ is indispensable, as shown in
Sec.~\ref{open_calibration}. \item The electromagnetic wave
penetrates deep into a non-metallic sample because -- in contrast
to a metal -- the real part of the dielectric permittivity is
positive. Field decay in the sample is only caused by the geometry
and determined by the electric contacts (and, only to some minor
degree, by the low absorption due to the imaginary part of $\e$).
In case of a 2.4~mm coaxial probe, the electric field strength
falls below 1~\%\ of its original value at the sample surface only
after penetrating more than 2.5~mm, for frequencies up to 5~GHz
and relative dielectric constant up to 50. With other words, the
penetration depth of the electric field is of the order of the
contact area dimensions (compare Fig.~\ref{sample}) and increases
rapidly with rising frequency and permittivity. Thus, in contrast
to the metals, the spatial field distribution which forms inside
an insulating semiconductor is significantly different from that
of a plane wave and it depends on frequency. Its knowledge is
essential to extract the complex conductivity $\sigma$ (or
permittivity $\e$) of such a sample from its complex impedance
$Z$. Integral equations, implying the accurate solution for the
electromagnetic field in the sample, cannot be directly solved for
$\e$. Hence, approximations are required or simplified models need
to be developed, with a limited range of validity for frequency
and permittivity.
\end{enumerate}

In the next three Sections we treat the problem of extracting the
material parameters of a semiconducting sample from the impedance
data. We first suggest in Sec.~\ref{static_model} a simple static
model which yields already a good approximation. In Sec.~\ref{LF}
we make use of the results by Levine and Papas \cite{LP} and Misra
\cite{M} to relate the complex impedance of a semiconducting
sample to its complex permittivity. In the following, the rigorous
electromagnetic field distribution from Sec.~\ref{LF} is used in
Sec.~\ref{open_calibration} to determine the frequency dependence
of the reflection coefficient $S^{\rm open}_{11}$ of the open
standard for the calibration step mentioned above.

\section{A static model for the relation between the sample
impedance and the complex conductivity} \label{static_model}

As a first approach to extract the conductivity of a low-loss
semiconducting sample from its impedance, we developed a simple
static model for the current distribution in a sample, based on
the following assumptions:
\begin{enumerate}
\item The response is local:
\begin{equation}
\vec{j}(\vec{r})=\sigma\cdot\vec{E}(\vec{r})\, \label{local},
\end{equation}
where $\vec{j}$ is the electric current density and $\vec{E}$ is
the electric field vector. This assumption implies that the
electric field does not vary significantly at the distances of the
mean free path $\ell$, which in the case of hopping transport is
the mean separation of the hopping partners. \item The dependence
of the electric field $\vec{E}(z, \rho)$ on the cylindrical
coordinates $z$ and $\rho$ can be accounted for separately. There
is no dependence on the angular coordinate $\varphi$ due to the
radial symmetry of the problem. \item Inside the coaxial line the
principal TEM mode is excited; thus, only the radial component of
the electric field $E(\rho)$ exists. The Gauss theorem yields
$E(\rho)=const./\rho$. With the voltage $U$ between the mask
contacts of radii $a$ and $b$ (Figs.~\ref{metal} and \ref{sample})
it follows:
\begin{eqnarray}
E(\rho)=\frac{U}{\ln\{b/a\}}\cdot\frac{1}{\rho}\, .
\end{eqnarray}
\item As far as the $z$ dependence of the electric field strength
is concerned, we assume that the field is concentrated at the
surface and gets weaker for further depth because the path length
for the corresponding current element d$I$ increases.

\begin{figure}
\centering
  \includegraphics*[width=0.75\linewidth]{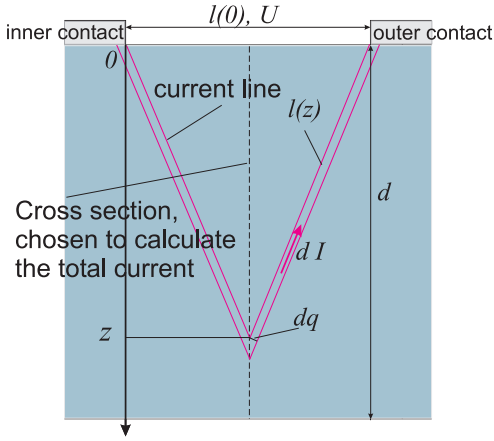}
  \caption{(color online) Geometry of the current distribution in a semiconducting sample of
thickness $d$, with metallic contacts at distance $l(0)=(b-a)$
(cf.\ Fig.~\ref{sample}) as assumed in the static model.}\label{model}
\end{figure}
In order to calculate the total current flowing through a sample,
we have chosen the cross section of the sample at the mid-distance
between the Al-contacts (Fig.~\ref{model}). The single current
line is approximated by a triangle shape with the apex at the
mid-distance cross section. Hence, we consider each infinitesimal
current line at its lowest point designated by the $z$ coordinate
and assume for the corresponding electric field $E(\rho,z)$ to be
reciprocally proportional to the length of the current line in
order to keep the voltage $U$ constant:
\begin{eqnarray}
E(\rho,z)=E(\rho)\cdot\frac{l(0)}{l(z)}\,, \label{E}
\end{eqnarray}
where $l(z)=l(0)\sqrt{1+[2z/l(0)]^2}$.
\end{enumerate}

Now that $E(\rho,z)$ is constructed, we can calculate the total
current $I$ flowing through a semiconducting sample using
Eqs.~(\ref{local})-(\ref{E}). The integral is taken over the entire
mid-distance cross section (see Fig.~\ref{model}) with the
infinitesimal element $\mbox{d}q=\mbox{d}z/\sqrt{1+[2z/l(0)]^2}$:
\begin{eqnarray}
\mbox{d}I(\rho,z)= j(\vec{r})\cdot\rho\,\mbox{d}\varphi\,\mbox{d}q&=&\sigma\cdot
E(\rho,z)\cdot\rho\,\mbox{d}\varphi\,\mbox{d}q\nonumber
\\ &=&\frac{\sigma\cdot U} {\ln\{b/a\}} \cdot\frac{\mbox{d}\varphi\,\mbox{d}z}{1+[2z/l(0)]^2}\;,
\nonumber\\
I=\int\limits_0^{2\pi}\int\limits_0^d\mbox{d}I(\rho,z)&=&
\frac{\sigma\cdot l(0)\,\pi
U}{\ln\{b/a\}}\arctan\left\{\frac{2d}{l(0)}\right\},\nonumber
\end{eqnarray}
where $l(0)=b-a$ (Fig.~\ref{model}). Thus, we have obtained a
relation between the complex impedance $Z=U/I$ at the sample
surface and the complex conductivity $\sigma$:
\begin{eqnarray}
\sigma=\frac{1}{Z}\cdot\frac{\ln\{b/a\}}{\pi (b-a)\arctan
\{2d/(b-a)\}}\; \label{model_sigma}.
\end{eqnarray}

In the limiting case of a thin conducting film there is a simple
geometrical relation between the conductivity and the impedance
because the $z$ dependence of the electric field and the boundary
effects can be neglected. This allows us to check the above
formula in the limit $d/l(0)\to 0$, where we recover the
Eq.~(\ref{film}).

\begin{figure}
\centering
  \includegraphics*[width=\linewidth]{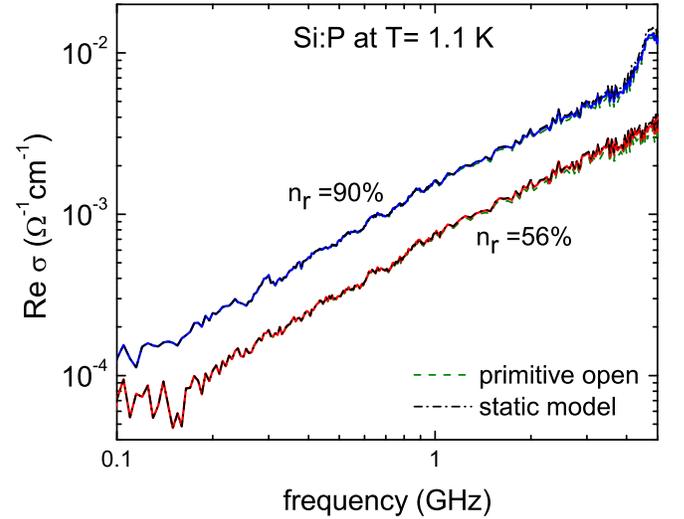}
  \caption{(color online) Frequency dependent conductivity of Si:P samples with
  relative donor concentration $n_r=0.56$ and 0.9 measured by a vector network analyzer at $T=1.1$~K.
  The solid lines represent the conductivity
obtained by the evaluation procedure described in Sec. \ref{LF}.
The dash-dotted lines result from evaluating the complex impedance
data using the formula (\ref{model_sigma}) of the static model.
The dashed lines stem from the primitive assumption $S^{\rm
open}_{11}=1$ for the open calibration standard. Concerning the
conductivity, all the three lines for either concentration $n_r$
are virtually identical, with only slight effects above 1~GHz.}
 \label{model_sig}
\end{figure}
\begin{figure}
\centering
  \includegraphics*[width=\linewidth]{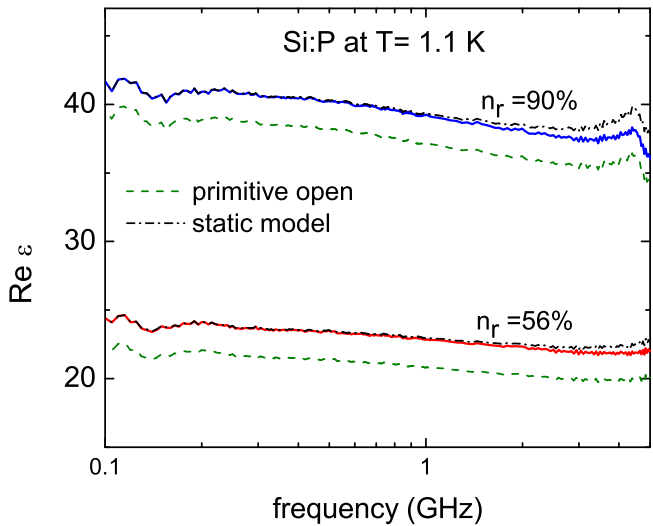}
  \caption{(color online) Typical permittivity
  spectra of Si:P-samples measured by a vector network analyzer at $T=1.1$~K.
  The solid lines correspond to the evaluation procedure described in Sec. \ref{LF}.
The dash-dotted lines result from evaluating the complex impedance
data using the formula (\ref{model_sigma}) of the static model.
The dashed lines stem from the primitive assumption $S^{\rm
open}_{11}=1$ for the open calibration standard.}\label{model_eps}
\end{figure}

The static model has been applied to analyze frequency dependent
impedance measurements of Si:P at low temperatures. As
demonstrated by the dash-dotted lines in Figs.~\ref{model_sig} and
\ref{model_eps}, the results of Eq.~(\ref{model_sigma}) agree very
well with the rigorous solution outlined in the following
Sections. Deviations can be noticed only above 1~GHz in the
dielectric constant.

\section{Relation between the complex impedance and the complex
permittivity of a semiconducting sample}\label{LF}
\subsection{General considerations}
A large variety of methods have been developed in the past to
extract the properties of low-loss and lossy dielectrics from a
reflection coefficient measurement
in the radio-frequency and microwave range. Simple ingenious models and
analytical solutions, that are valid in a limited parameter range,
have been suggested; time-consuming but arbitrarily precise
numerical approaches have been treated depending on the specific
practical goals. A comprehensive list of references is available
in review articles like Ref.~\onlinecite{PM}.

In studies of soft and liquid materials, the coaxial probe was
frequently modelled as an equivalent circuit consisting of several
fringe-field capacitors in the lumped-element approach
\cite{Stuchlies,Grant,Clark1,Clark2}. In Ref.~\onlinecite{Grant} a
comprehensive, detailed and critical revision of this method can
be found. The most striking point is the strong dependence of the
model capacitances on the permittivity of the material that
terminates the coaxial line, thus the approach is limited to
specimens with dielectric properties close to those of the
reference materials available.

Here we consider a convenient analytical way to extract the
complex conductivity $\sigma$ from the sample impedance $Z$ based
on the works of Levine and Papas \cite{LP} and Misra \cite{M}. The
method is valid at least up to 5~GHz for the 2.4~mm probe and
relative dielectric constants up to 50; an extension to higher
frequencies is possible with certain numerical procedures added.
As an intermediate result, there is an integral expression for the
sample admittance $Y=1/Z$ as a function of the material dielectric
function $\e$, which is well suited to determine the frequency
dependence of the open calibration standard in a closed manner
(Secs.~\ref{problem} and \ref{open_calibration}). The theoretical
expressions for the electromagnetic field on both sides of the
sensor aperture\cite{LP} are rewritten for the case of a medium
with an arbitrary complex permittivity in the sample half-space.
For the parameter range considered here \cite{higher_modes}, we
regard the variational principle applied by Levine and Papas as
preferable to the precise but time-consuming numerical method of
point-matching proposed by Mosig \cite{Mosig,Grant}.

\subsection{Solution in form of an integral equation}

\begin{figure}
  \includegraphics*[width=0.9\linewidth]{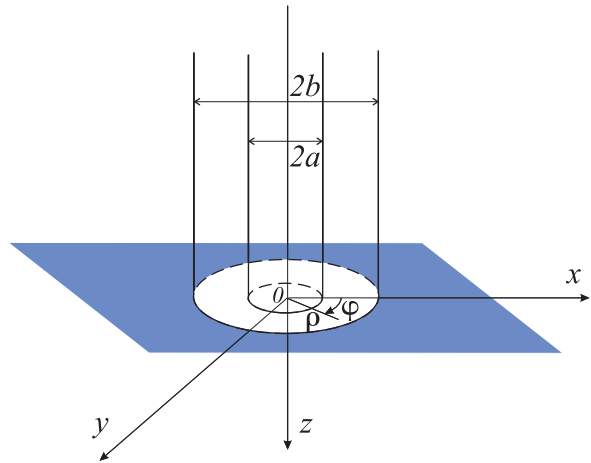}
  \caption{(color online) The plane $z=0$ constitutes the interface between the coaxial wave
guide ($z<0$) and the sample space ($z>0$). It serves as a
reference plane, at which the reflection coefficient $S_{11}$, the
impedance $Z$ and the admittance $Y$ of the sample are
defined.}\label{flange}
\end{figure}
In the following, the coaxial wave guide with a center conductor
of radius $a$ and an outer conductor of radius $b$ is terminated
by an infinite-plane conducting flange at $z=0$
(Fig.~\ref{flange}). Choosing the dimensions $a$ and $b$ of the
coaxial line to be small enough, the assumption of a single
propagating mode (the principal TEM mode) in the coaxial region is
justified in the covered frequency range. This system is amenable
to a detailed theoretical analysis \cite{LP} that yields the
electromagnetic field distribution in the half-space $z>0$ and a
relation between the aperture admittance $Y$ (current-to-voltage
ratio at $z=0$) and the complex wave vector $\vec{k}$ of the free
space $z>0$.

We may assume an insulating sample to fill the half-space $z>0$ by
choosing its finite dimensions to be large enough for the
electromagnetic field strength so that the sample boundaries are
negligible (cf. Sec.~\ref{problem}). The results for the free
space thus transform into a relation between the admittance $Y$
measured at the sample surface $z=0$ and its complex dielectric
function $\e$ \cite{full_wave}(here, a harmonic time dependence
$\exp\{-i\omega t\}$ is assumed for the electromagnetic
field\cite{notation}):
\begin{eqnarray}
\frac{Y}{Y_0}=\frac{-ik^2}{\pi
k_c\ln\{b/a\}}\int\limits_a^b\int\limits_a^b\int\limits_0^\pi
\cos\varphi\, \frac{\exp\{ik r\}}{r} \,\mbox{d}\varphi\,
\mbox{d}\rho\, \mbox{d}\rho^{\,\prime} \label{LPM},
\end{eqnarray}
where
\begin{eqnarray}
&&r=(\rho^2+\rho^{\,\prime2}-2\rho
\rho^{\,\prime}\cos\varphi)^{1/2}\,,\\
&&k^2=\omega^2\mu_0\e_0\e\,\label{k}
\end{eqnarray}
and $k_c$ is the wave vector of the coaxial line with the
characteristic admittance \cite{Pozar}
\begin{eqnarray}
Y_0=1/Z_0=2\pi\cdot\left(\sqrt{\frac{\mu_0}{\e_0\e_c}}\ln\{b/a\}\right)^{-1}\,.
\label{Y0}
\end{eqnarray}
The validity of the variational approximation (\ref{LPM}) has been
proven in Ref.~\onlinecite{LP} in the parameter range $0<ka\leq 2$
and $1.57\leq b/a \leq 4$ by comparison with experimental results.
To obtain low-temperature microwave data on Si:P with widely
varying phosphorus concentration $n$, we employ a coaxial probe of
dimensions $2b=1.75$~mm, $2a=0.6$~mm. The maximum frequency range
spans from 45~MHz to 40~GHz (limited by source and test set of the
network analyzer HP 8510) and the relative dielectric constants
$\eps$ reach from 18 till 50. This corresponds to $0.001 \leq
ka\leq 1.76$ and $b/a=2.9$ and lies within the tested parameter
range.

\subsection{Solution of the inverse problem}

The inverse problem of extracting $\e$ from the measured impedance
$Z$ using Eq.~(\ref{LPM}) has been solved in the quasi-static
approximation by Misra \cite{M}. For low frequencies, the
exponential function in Eq.~(\ref{LPM}) can be approximated by the
first four terms of its series expansion:
\begin{eqnarray}
Y\approx\frac{-i2\omega\e_0\e}{[\ln\{b/a\}]^2}\int\limits_a^b\int\limits_a^b
\int\limits_0^\pi\left[\frac{\cos\varphi}{r}+ik\cos\varphi\right.
\nonumber\\
\left.-\frac{k^2r}{2}\,\cos\varphi-i\,\frac{k^3r^2}{6}\,\cos\varphi\right]\mbox{d}\varphi\,
\mbox{d}\rho\, \mbox{d}\rho^{\,\prime}\,. \label{expansion}
\end{eqnarray}
The second term of Eq.~(\ref{expansion}) vanishes upon
integration, and the last one is readily integrated; the integrals
corresponding to the first and the third terms need to be
numerically evaluated:
\begin{equation}
Y\approx\frac{-i2\omega\e_0\e}{(\ln\{b/a\})^2}
\left[I_1-\frac{k^2I_3}{2}\right]+
\,\frac{k^3\pi\omega\e_0\e}{12}\,\left[\frac{b^2-a^2}{\ln\{b/a\}}\right]^2
, \label{radiation}
\end{equation}
where
\begin{eqnarray}
I_1=\int\limits_a^b\int\limits_a^b\int\limits_0^\pi\frac{\cos\varphi}{(\rho^2+\rho^{\,\prime2}-2\rho\rho^{\,\prime}\cos\varphi)^{1/2}}\mbox{d}\varphi\,
\mbox{d}\rho\, \mbox{d}\rho^{\,\prime}\nonumber
\end{eqnarray}
and
\begin{eqnarray}
I_3=\int\limits_a^b\int\limits_a^b\int\limits_0^\pi\cos\varphi\,(\rho^2+\rho^{\,\prime2}-2\rho\rho^{\,\prime}\cos\varphi)^{1/2}\mbox{d}\varphi\,
\mbox{d}\rho\, \mbox{d}\rho^{\,\prime}\,.\nonumber
\end{eqnarray}
In our special case the relative contribution of the last term in
Eq.~(\ref{radiation}) to $Y$ is below 1\%\ up to 5~GHz
 and the formula thus reduces to a quadratic
equation for $\e$:
\begin{eqnarray}
\frac{1}{Z}=Y=\frac{-i2\omega\e_0\e}{(\ln\{b/a\})^2}\left[I_1-\frac{\omega^2\mu_0
I_3\e_0\e}{2}\right]\, .  \label{qequation}
\end{eqnarray}\\
The values of the geometrical integrals for the special case of
the coaxial probe with the inner and outer conductor diameters
$2a=0.6$ mm and $2b=1.75$ mm are listed in Tab.~\ref{integrals}.
\begin{table}\caption{Geometrical integrals for the coaxial probe dimensions $2a=0.6$ mm and $2b=1.75$ mm}\label{integrals}
\begin{tabular}{lllll}
&&&&\\\hline
     $I_1$, mm\qquad\qquad & $I_3$, mm$^3$\qquad\qquad & $I_4$, mm$^4$\qquad\qquad & $I_5$, mm$^5$\\\hline
     0.9084 & -0.2100 & -$(\pi/4)\cdot$0.4001\qquad\qquad & -0.4047\\\hline
\end{tabular}
\end{table}
It should be mentioned that the integrand of $I_1$ diverges at
$\rho=\rho^{\,\prime},\, \varphi=0$. That integral was numerically
evaluated as the limit of a series of integrals $\{I_{1,\,n}\}$
which lower bounds $\varphi_n$ converge to $\varphi=0$.

\section{Open calibration standard} \label{open_calibration}

The frequency dependence of the open calibration standard with a
known dielectric function $\e$ can be obtained as follows. The
expression (\ref{LPM}) of the admittance $Y$ as a function of $\e$
describes the open standard admittance correctly as long as the
effect of the finite sample dimensions is negligible. Using a
teflon block of the form shown in Fig.~\ref{sample} and assuming
its dielectric function to be $\e=2.03(1+i\,0.0002)$ in the GHz
frequency range \cite{teflon}, the maximum electric field strength
at the depth of 2 mm turns out to be far below 0.01 of its value
at the sample surface for frequencies up to 10 GHz, so that the
secondary reflections at the back side of the open standard can be
neglected here.

In order to obtain a closed expression $Y^{\rm
open}(\omega)$ from the integral equation (\ref{LPM}), the series
expansion of the exponential function can be used as in the
previous section. In contrast to the inverse problem discussed in
Sec.\ \ref{LF}, there is no need to spare at the accuracy
truncating the series early here. The relative contribution of the
subsequent term being far below 10$^{-4}$ up to 10~GHz, the
ultimate expression we use is:
\begin{equation}
Y^{\rm open}\approx\frac{-i2\omega\e_0\e}{(\ln\{b/a\})^2}
\left[I_1-\frac{1}{2}k^2I_3-\frac{i}{6}k^3I_4+\frac{1}{24}k^4I_5\right],\;\;\label{open_expansion}
\end{equation}
where
\begin{eqnarray}
I_4&=&-\frac{\pi}{4}(b^2-a^2)^2\,,\nonumber\\
I_5&=&\int\limits_a^b\int\limits_a^b\int\limits_0^\pi\cos\varphi\,(\rho^2+\rho^{\,\prime2}-2\rho\rho^{\,\prime}\cos\varphi)^{3/2}\mbox{d}\varphi\,
\mbox{d}\rho\, \mbox{d}\rho^{\,\prime}.\nonumber
\end{eqnarray}
and $k$ is defined in Eq.~(\ref{k}). The values of the geometrical
integrals for the 2.4~mm coaxial probe are listed in Table~\ref{integrals}.

The frequency dependent reflection coefficient $S^{\rm open}_{11}$
of the open calibration standard follows using Eq.~(\ref{Z-S}):
\begin{eqnarray}
S^{\rm open}_{11}=\frac{Z^{\rm open}-Z_0}{Z^{\rm open}+Z_0} =
\frac{Y_0-Y^{\rm open}}{Y_0+Y^{\rm open}}\, \label{Y-S}.
\end{eqnarray}
The effect of the frequency dependence of $S^{\rm open}_{11}$ on
the conductivity $\sig$ and permittivity $\eps$ spectra compared
to the dc assumption $S^{\rm open}_{11}=1$ is demonstrated in
Figs.~\ref{model_sig} and \ref{model_eps} on the example of two
Si:P-samples with donor concentration $n/n_c$ of 0.56 and 0.9
relative to the concentration value at the metal-insulator
transition, $n_c=3.5\times 10^{18}$ cm$^{-3}$. For samples with
larger dielectric constant $\eps$ and higher losses $\sig$, as
$n/n_c$ rises, the influence of this correction slightly
decreases. This is also what one would expect, when the electric
properties of the material under investigation approach the
metallic characteristics.

\section{Application to the hopping transport in Si:P}
\label{application}
\subsection{Dynamical conductivity}
With the method described in the previous parts, Secs.~\ref{LF}
and \ref{open_calibration}, we study the frequency-dependent
hopping transport in Si:P in order to explore the influence of
electronic correlations \cite{Ritz}. At concentrations of
phosphorus in silicon below the critical value of $n_c=3.5\times
10^{18}$ cm$^{-3}$, the donor electron states are strongly
localized due to disorder in Anderson sense
\cite{Anderson,ESbook}. Since some degree of compensation by
impurities of the opposite type is inevitable, charge transport at
low excitation energies is by variable-range hopping between the
donor sites, randomly distributed in space \cite{ESbook,Mott}.
Thus, theoretical
  models for a disordered system with electron-electron interaction are appropriate
  to interpret the electric conductivity spectra \cite{ES}. The main issue we address is that of power
laws of the frequency-dependent conductivity at zero temperature:
\begin{eqnarray}
\quad \sig(\omega)\sim\omega^\alpha\,, \quad
\sigma=\sig+i\sigma_2\,.\label{power_law}
\end{eqnarray}

From the theory of resonant photon absorption by pairs of states,
one of which is occupied by an electron and the other one is
empty, distinct limiting results are known for the conductivity
power law depending on which of the relevant energy scales of the
problem dominates over the others:

Taking into account the Coulomb repulsion $\WW$ if both states in
a pair would be occupied by an electron ($\rhop$ is the most
probable hopping distance), Shklovskii and Efros derived
$\sig(\omega)$ to be a sub-linear function of frequency, as long
as the Coulomb interaction term dominates over the photon energy
\cite{ES}. At higher frequencies, in the opposite limit, the
sub-quadratic behavior known from Mott \cite{Mott} for
non-interacting electrons is recovered. In addition, it is known
that due to electronic correlations an area of reduced density of
states is formed around the Fermi level, the so-called Coulomb gap
$\Delta$. For the conductivity of interacting electrons where the
Coulomb term $\WW$ dominates over the photon energy but falls
inside the Coulomb gap $\Delta$, the reduction of the density of
states leads to a stronger, slightly super-linear power law
\cite{ES}. In order to gain some insight into the effects of
electronic correlations, it is required to extract thoroughly the
frequency-dependent conductivity and the related power-laws over a
wide spectral range for a variety of doping concentrations.

\begin{figure}
  \includegraphics*[width=\linewidth]{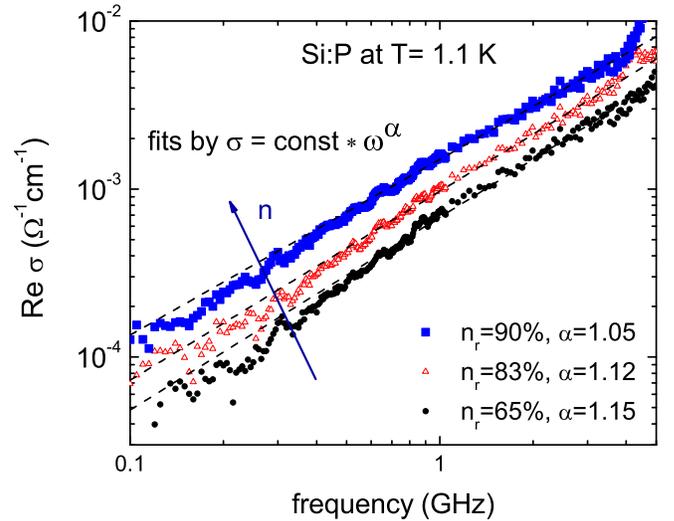}
  \caption{(color online)  Typical spectra of the measured real part of the
  conductivity for Si:P with relative donor concentrations $n/n_c$ of 0.65, 0.83 and 0.90.
  The dashed lines are the fits by a two-parameter power law function.}\label{sigma_1K}
\end{figure}
In Fig.~\ref{sigma_1K} the measured real part of the
frequency-dependent conductivity is plotted on a log-log scale to
identify the power law. The fits by a two-parameter function
$\sig(\omega)=\mathrm{const.}\cdot\omega^\alpha$ are shown by the
dashed lines. The frequency dependence of the conductivity clearly
follows a super-linear power law in the whole doping range, where
the exponent decreases with doping \cite{Ritz}. From this we
infer, that hopping transport takes place deep inside the Coulomb
gap $\Delta$. A super-linear conductivity power law was previously
observed in Si:As and Si:P by Castner and collaborators
\cite{DC,MC} using resonator techniques at certain frequency
within the range of the present work.  Our results are in accord
with the measurements on similar samples at higher frequencies
(30~GHz to 3~THz) using optical techniques \cite{Hering1,Hering2}.
In contrast, a sub-linear frequency dependence in the zero-phonon
regime has been reported by Lee and Stutzmann \cite{Lee} based on
experiments on Si:B in the microwave range and by Helgren et al.
\cite{Helgren} using quasi-optical experiments.

\begin{figure}
\centering
  \includegraphics*[width=\linewidth]{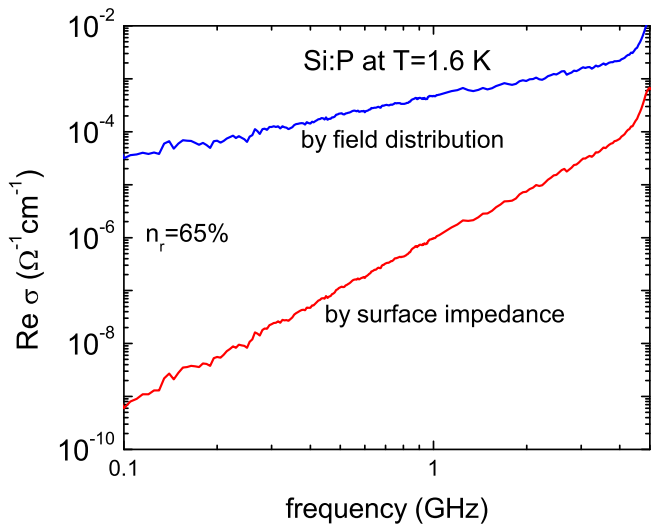}
  \caption{(color online) Frequency-dependent conductivity  of a thick
  Si:P-sample with relative donor concentration of 65\%\ measured at 1.6 K.
  In the first case (red line) the evaluation of the impedance data is done
  by the surface impedance concept (Sec.~\ref{metallic}). Using the correct
  field distribution (Sec.~\ref{LF}), a significantly different frequency dependence
  (blue line) is obtained from the same raw data.}
\label{evaluation}
\end{figure}
 As demonstrated in Fig.~\ref{evaluation}, the surface
impedance approach, mentioned in Sec.~\ref{metallic}, yields a
wrong (i.e.\ too strong) frequency dependence of the conductivity
$\sig(\omega)$ for insulators. The evaluation of the impedance
spectrum of a Si:P-sample is shown using the surface impedance
formula (\ref{sigma-Zs}) in comparison to the solution of the
equation (\ref{qequation}). The latter method yields a
conductivity power law of approximately one, as expected from the
theory outlined above.

\subsection{Dielectric function}
It is an additional advantage of a phase sensitive measurement to
gain the dielectric function $\eps$ from the imaginary part of the
complex conductivity \cite{Drs}:
\begin{eqnarray}
\e=\e_1+ i\,\e_2= 1+ i \frac{\sigma}{\e_0\omega}\, ,
\label{epsilon-sigma}
\end{eqnarray}
We denote by $\e$ the full complex dielectric function of Si:P,
 relative to the free space
permittivity $\e_0$. As the metal-insulator transition is
approached upon doping $n$, the localization radius diverges
\cite{ESbook}. As a consequence, the electronic contribution to
the dielectric function is also expected to diverge following a
power law when the metal-insulator transition is approached
\cite{ESepsilon}:
\begin{eqnarray}
\quad \eps-\e_{\rm Si}\sim|1-n/n_c|^{-\zeta}\,, \label{scaling}
\end{eqnarray}
where $\e_{\rm Si}=11.7$ is the dielectric constant of the host
material Si.

\begin{figure}
  \includegraphics*[width=\linewidth]{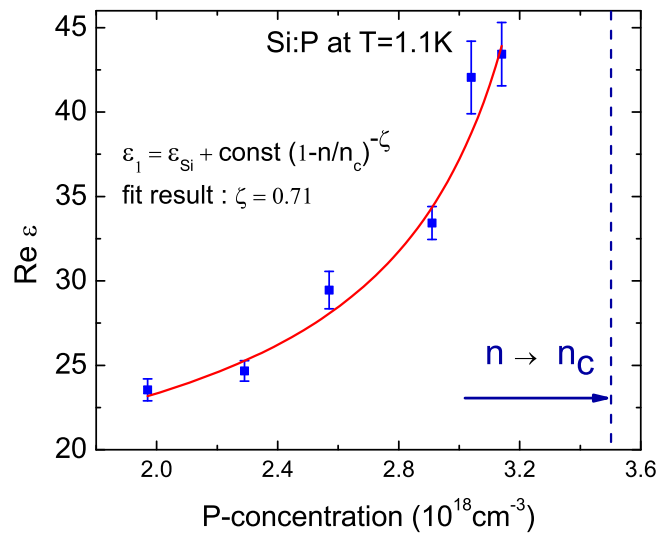}
  \caption{(color online) Doping dependence of the low-temperature dielectric constant
  of Si:P. The solid line is the fit by
  the power law function (\ref{scaling}) for the electronic dielectric
  function $\eps-\e_{\rm Si}$.}\label{fig:epsilon}
\end{figure}
From our experiments \cite{Ritz} we find that the dielectric
function $\eps$ is independent of frequency in the range from 50
MHz to 10 GHz, taking the measurement uncertainty into account
(Fig.~\ref{model_eps}). A fit with the function (\ref{scaling})
results in an exponent $\zeta=$0.71, as shown in
Fig.~\ref{fig:epsilon}.

In the framework of the effective medium approximation $\zeta=1$
is expected  \cite{ESepsilon}. From the quasi-optical experiments
on Si:P different results are reported. Helgren et al.
\cite{Helgren} observe a similar dependence of the values of the
dielectric constant on the donor concentration (though uniformly
shifted to lower values by 8). Hering et al. \cite{Hering2} have
observed values of $\eps$ as we have, but with a much stronger
donor concentration dependence of the dielectric constant,
resulting in a much higher exponent $\zeta=1.68$. It is obvious
that this enormous discrepancy  calls for further experiments
which are more accurate as far as this analysis is concerned.

\section{Conclusions}
We have thoroughly analyzed the problem of extracting the
electrical conductivity and permittivity from the complex
impedance measured in the microwave range using a network
analyzer. While for thin metallic films and bulk metals simple
relations are readily available, special care has to be taken in
the case of semiconductors and insulators where the electric field
penetrates and decays over an appreciable distance. Already a
static model with an approximate field configuration leads to
reasonable results. Eventually we present a rigorous solution of
the problem with basically no restricting assumptions. The
advanced analysis is applied to the broadband impedance
measurements of Si:P with different doping concentrations and
spanning a wide range of frequency. The findings can now be
compared to the theory and yield important insight into the
effects of electronic correlations on the hopping transport at low
temperature.

\section*{Acknowledgements}
The work was partially funded by the Deutsche
Forschungsgemeinschaft (DFG). ER would like to acknowledge a
scholarship by the Landesgraduierten-Programm of
Baden-W\"urttemberg.


\begin{thebibliography}{}
\bibitem{simplified} In literature (e.g. Ref.~\onlinecite{Anlage}) a simplified formula can be found, that
follows from the equation (\ref{sigma-Zs}) if $\e_0$ is neglected
and the sign of the imaginary part of $Z_S$ is defined in a
different way according to the complementary sign notation
$\exp\{i(\omega t-\vec{k}\vec{x})\}$ in the exponent of the
Fourier-transformed of the electromagnetic fields (cf.
Ref.~\onlinecite{notation}).
\bibitem{higher_modes} The variational approximation applied by Levine and Papas \cite{LP} leads
to a vanishing contribution of the integral terms corresponding to
the evanescent higher order terms of the electromagnetic wave
inside the coaxial sensor. The parameter range, where the
corresponding uncertainty in the resulting capacitances is below
1\%, as proved by Levine and Papas, includes our range of
frequency and dielectric function. Moreover, Misra \cite{M} has
critically compared the method with the results obtained by Mosig
\cite{Mosig} under consideration of higher order modes, yielding a
parameter range where the variational approximation is justified.
\bibitem{full_wave}For effectively smaller samples, a full wave treatment considered
by Brom and collaborators \cite{B} is necessary, which takes
secondary reflections at the sample back into account.
\bibitem{notation} We hold to the convention
$\exp\{i(\vec{k}\vec{x}-\omega t)\}$ for the Fourier-transformed
of the electromagnetic field in accordance with
Refs.~\onlinecite{Drs,LP}. The opposite sign convention is used in
Refs.~\onlinecite{B,M} and by the network analyzer.
\end{thebibliography}
\end{document}